\documentclass[12pt]{amsart}
\usepackage{amsmath}
\usepackage{amssymb}
\usepackage[latin1]{inputenc}
\usepackage[T1]{fontenc}
\newcommand{\N}{\mathbb{N}}
\newcommand{\R}{\mathbb{R}}
\newcommand{\be}{\begin{equation}}
\newcommand{\eeq}{\end{equation}}
\newcommand{\bet}{\begin{equation*}}
\newcommand{\eeqt}{\end{equation*}}
\newcommand{\bea}{\begin{eqnarray}}
\newcommand{\eeqa}{\end{eqnarray}}
\newcommand{\beat}{\begin{eqnarray*}}
\newcommand{\eeqat}{\end{eqnarray*}}

\newcommand{\hil}{\mathcal{H}}

\newcommand{\br}{\mathcal{B}(\R)}
\newcommand{\lh}{\mathcal{L(H)}}
\newcommand{\sfq}{\mathsf{Q}}
\newcommand{\sfp}{\mathsf{P}}
 
\newcommand{\E}{\mathsf{E}}
\newcommand{\G}{\mathsf{G}}
\newcommand{\Eh}{\mathsf{E}_{\rm ht}}

\newcommand{\tr}[1]{\mathrm{tr}\left[ {#1} \right]} 
\begin{document}
\title{On the complementarity of the quadrature observables}

\author{Pekka Lahti}
\address{Turku Centre for Quantum Physics, Department of Physics and Astronomy, University of Turku, 20014 Turku, Finland}
\email{pekka.lahti@utu.fi}
\author{Juha-Pekka Pellonp\"a\"a}
\address{Turku Centre for Quantum Physics, Department of Physics and Astronomy, University of Turku, 20014 Turku, Finland}
\email{juha-pekka.pellonpaa@utu.fi}

\begin{abstract}
In this paper we investigate the coupling properties of pairs of quadrature observables, showing that, apart from the Weyl relation, they share
the same coupling properties as the position-momentum pair. 
In particular, they are complementary.
We determine the marginal observables of a covariant phase space observable with
respect to an arbitrary rotated reference frame, and observe that these marginal observables are unsharp quadrature observables. The related
distributions constitute the Radon tranform of a phase space distribution of the covariant phase space observable. Since the quadrature distributions 
are the Radon transform of the Wigner function of a state, we also exhibit the relation between the quadrature observables and the tomography observable,
and show how to construct the phase space observable from the quadrature observables. Finally, we give a method to measure together with a single measurement scheme any
complementary pair of  quadrature observables.\newline

\

\noindent
Dedicated to Peter Mittelstaedt in honour of his eightieth birthday.
\end{abstract}
\maketitle

\section{Introduction}
The notion of complementarity was introduced to the vocabulary of quantum physics by Niels Bohr in his famous Como lecture of 1927 as a key 
to the understanding of quantum phenomena in terms of classical concepts \cite{Bohr1928}. One of the most explicit uses of the "traffic rules" of Bohr
was in his 1935 paper \cite{Bohr1935}, where he argued that position and momentum of a particle are complementary quantities in the sense that all 
the experimental
arrangements allowing their unambiguous operational definitions are mutually exclusive but they both are needed for a full description of
the situation.

In addition to the position-momentum pair, energy-time, path-interference, number (action)-phase, spin-phase, or spin components,
 are  frequently occuring examples of pairs of complementary observables. Moreover, complementary modes of description, like the use
of complementary bases, or  the past and the future state determinations of the system, are often discussed cases.
For an overview of various aspcets of this notion
we refer to
\cite{BL1995,OQP}.

In this paper we investigate the properties of pairs of quadrature observables in close analogy to the position-momentum case, reviewed in section \ref{qp}. We show in 
section \ref{quadrature} that --
apart from the Weyl relation -- they share all the  coupling propeties of position and momentum, which reflect the strong incompatibility
of these observables. In sections~\ref{phasespace} and \ref{tomography} we exhibit the connection of the quadrature observables to the covariant
phase space observables and the tomography observable, respectively. In the concluding section~\ref{measuring} we demonstrate, that though
any pair of quadrature observables is complementary in the sense that none of their measurements can be combined into a joint 
measurement of theirs, there are single measurement schemes which allow one to determine the measurement outcome distributions of the given
quadrature pair for a large class of states of the system.

\section{The pair $(Q,P)$}\label{qp}
Let ${\mathcal H}=L^2(\R)$ be the usual $L^2$-function space on $\R$ spanned by Hermite functions $h_n$, $n\in\N=\{0,1,...\}$.
Consider the selfadjoint position operator $Q$ on the Hilbert space ${\mathcal H}$ of a quantum object in one dimension,
and let $\sfq$ be its spectral measure, so that $\sfq(X)$ is the multiplication with the characteristic function $\chi_X$ of the (Borel)
set $X\in\br$. Let $F$ be the unitary Fourier-Plancherel  operator on $L^2(\R)$ so that $P=F^{-1}QF$ is the selfadjoint
momentum operator $-id/dx$, with the spectral measure $\sfp=F^{-1}\sfq F$.  For any state $\rho$ (positive trace-1 operator)
we let $\rho^Q$ and $\rho^P$ denote the densities of the probability measures $X\mapsto p^{\sfq}_\rho(X)=\tr{\rho\sfq(X)}$
and $Y\mapsto p^{\sfp}_\rho(X)=\tr{\rho\sfp(Y)}$ with respect to the Lebesgue measure.

The unitary groups 
$\{e^{-iqP}\,|\, q\in\R\}$ and $\{e^{ipQ}\,|\,p\in\R\}$ 
of $Q$ and $P$ fulfill the Weyl relation
\begin{equation}\label{Weyl}
e^{-iqP}e^{ipQ}=e^{-iqp}e^{ipQ}e^{-iqP},
\end{equation}
 and, modulo unitary equivalence, the pair $(Q,P)$ is uniquely determined by this relation \cite{vN1931}.
In addition, the pair $(Q,P)$ has the following well-known  coupling properties:

\

\begin{itemize}
\item[a)] $QP-PQ=iI$ (on the (dense) domain of the commutator);
\item[b)] $\inf\{{\rm Var}(\rho^Q)\cdot{\rm Var}(\rho^P)\,|\, \rho\ {\rm a\ state}\,\}=\frac 14 >0$;
\item[c)] ${\rm com}(Q,P)=\{\psi\in L^2(\R)\,|\, \sfq(X)\sfp(Y)\psi=\sfp(Y)\sfq(X)\psi\ {\rm for\ all}\ X,\,Y\in\br\} =\{0\}$;
\item[d)] $\sfq(X)\land\sfp(Y)= \sfq(X)\land\sfp(\R\setminus Y)= \sfq(\R\setminus X)\land\sfp(Y)= 0$ for all bounded $X,\,Y\in\br$;
\item[e)] $\tr{\sfq(X)\sfp(Y)}=\frac{1}{2\pi}\lambda(X)\lambda(Y)$ for all bounded $X,Y\in\br$, with $\lambda(X)$ being
the Lebesgue measure of $X$.
\end{itemize}

\

\noindent
In addition to the Weyl relation, all the properties $a)$ through $e)$ reflect extreme incompatibility of position and momentum observables.
The commutation relation $a)$ as well as the preparation uncertainty relation $b)$ belong to the basic arsenal of quantum mechanics and 
need no further comments here.
The property $c)$ expresses the fact that for no state $\rho$
the map $(X,Y)\mapsto \tr{\rho\sfq(X)\land\sfp(Y)}$
extends to a probability measure on $\mathcal B(\R^2)$ \cite{Ylinen1985}.
On the other hand,   relations $d)$ and $e)$ have been taken to describe the complementarity of these observables
in the sense of lack of any joint measurements, see, for instance,
\cite[III.8..2]{QTM},
\cite[IV.2.3]{OQP}, or, as a kind of generalization  of the "complementary bases" of the finite dimensional case
 \cite{Accardi1984,Kraus1987}. It is, perhaps, well-known, and  will also be shown below that
none of these five properties $a)-e)$ is sufficient to determine the pair $(Q,P)$ to be the Weyl pair, for an explicit proof of the case $e)$,
 see, e.g.\ \cite{GianniVSV2002}. 

It is also well-known that the pair $(Q,P)$ is informationally incomplete: the measurement outcome statistics $\rho^\sfq,\rho^\sfp$ 
of these observables do not
suffice, in general, to determine the state $\rho$ of the system.\footnote{One of the first examples demonstrating this fact is reported
in \cite{Reichenbach} and is  due to V.\ Bargmann.}
In the words of C.F. von Weizäcker \cite{CFvW85}, this is a reflection of the surplus of information coded in the quantum notion of state,
when compared with the classical one. 

\section{The pair $(Q,Q_\theta)$}\label{quadrature}

Position $Q$ and momentum $P$ can be obtained in a smooth way from each other. Indeed, let $U_\theta=e^{i\theta H}$,
$\theta\in\R$, be the unitary operator defined by the oscillator operator $H=\frac 12(Q^2+P^2)$, and define 
$Q_\theta=U_\theta QU_\theta^*$, so that $Q_\theta$ is the quadrature operator, 
with the spectral measure $\sfq_\theta=U_\theta\sfq U_\theta^*$.
Clearly $Q_0=Q$ and $Q_{\pi/{2}}=P$ (since $F=U_{-\pi/2}$); in fact, $Q_\theta =Q\cos\theta + P\sin\theta$. To study the coupling properties of any two
quadratures $(Q_\alpha,Q_\beta)$, it is sufficient to consider the pair $(Q,Q_\theta)$, $\theta\in[0,2\pi)$,
since for any pair $(Q_\alpha,Q_\beta)$ one finds a unitary operator $U:=U_\alpha$ such that
$Q_\alpha=U Q U^*$ and $Q_\beta =U Q_{\beta-\alpha} U^*$, that is, the pair $(Q_\alpha,Q_\beta)$ is unitarily equivalent to the pair $(Q,Q_\theta)$ where $\theta=\beta-\alpha$.

Using the  operator relation 
$$e^{iy(Q\cos\theta +P \sin\theta)}=e^{iyQ\cos\theta}e^{iyP\sin\theta}e^{i y^2(\cos\theta\,\sin\theta)/2}$$
together with the Weyl relation
 one checks that the unitary operators $e^{ixQ}$ and $e^{iyQ_\theta}$
fulfill the Weyl relation exactly when $\theta=\pi/2$, that is, $Q_\theta =P$.

Clearly, for any pair of quadratures $(Q,Q_\theta)$,
$$
QQ_\theta-Q_\theta Q= i\sin\theta\, I
$$
on the dense domain of the commutator. 
Denote $b:=\sin\theta$ and assume that $b\ne 0$.
 Since $(\frac 1b Q,Q_\theta)$ is a Weyl pair,  
and the spectral projections of $\frac 1b Q$ are of the form $\sfq(bX)$, $X\in\br$, one notes that all 
the coupling properties $b)-e)$  hold for the pair $(Q,Q_\theta)$, as well.
In particular,  for any $\theta\not\in \{0,\pi\}$,
\begin{equation}\label{UR}
\inf\{{\rm Var}(\rho^{\sfq})\cdot{\rm Var}(\rho^{\sfq_\theta})\,|\, \rho \ {\rm a\ state}\,\}=\frac {\sin^2\theta}4 >0,
\end{equation}
 the quadratures  $\sfq$ and $\sfq_\theta$ are totally noncommutative\footnote{We recall from  \cite{DP} that condition (\ref{UR})
alone implies that ${\rm com}(Q,Q_\theta)=\{0\}$.},
they do have no joint probability of the form $(X,Y)\mapsto\tr{\rho\sfq(X)\land\sfq_\theta(Y)}$, they are complementary
in the sense of $d)$  and they satisfy  the trace formula $e)$.

As already pointed out, the pair $(Q,P)$ is informationally incomplete. 
Clearly, the same is true for any pair $(Q,Q_\theta)$.
These  pairs can, however,  be completed adding further quadratures. Indeed,
any set $\{\sfq_\theta\,|\,\theta\in S\}$, where $S\subset[0,\pi)$ is a dense set, is  informationally complete, and thus allows 
state determination on the basis of the statistics $\rho^{\sfq_\theta},\theta\in S$, see, e.g. \cite{Gianni2000,KLP2008}. 
The other  method to complete the pair $(Q,P)$ is to replace it by a coexistent pair 
$(\mu*\sfq,\nu*\sfp)$ of unsharp position and momentum observables such that their joint observable is informationally complete 
\cite{AliDoeb,AliPru}. 
We recall that, for instance, $\mu*\sfq$ is the normalized positive operator measure ({\small POM})  defined by the convolution of the probability measures 
$\mu$ and $p^\sfq_\rho$, 
$$
\tr{\rho (\mu*\sfq)(X)}=(\mu* p^\sfq_\rho)(X)=\int_\R\mu(X-q)\, dp^\sfq_\rho(q)=\int_\R\mu(X-q)\rho^\sfq(q)dq
$$
where $\rho$ is a state. 
As will be seen below, these two approaches are  closely related with each other.

\section{The pairs $(Q,Q_\theta)$ and the phase space pom $\G_K$}\label{phasespace}
Due to the noncommutativity of the pair  $(Q,Q_\theta)$, there is no normalized positive operator measure ({\small POM}) $\E:\mathcal B(\R^2)\to\lh$ 
which would have both $\sfq$ and $\sfq_\theta$ as the marginal observables \cite[Thm IV.\ 1.3.1.]{Ludwig83}.
On the other hand, the unsharp pair $(\mu*\sfq,\nu*\sfp)$ has joint observable exactly when the probability measures $\mu$ and $\nu$ have
Fourier related densities, in which case $(\mu*\sfq,\nu*\sfp)$ are the (Cartesian) marginal observables of a covariant phase space observable $\G_K$
generated by a
positive trace-1 operator $K$ on $\hil$  \cite{CHT2005}. 
We recall that $\G_K$
is defined by the operator density $(q,p)\mapsto W(q,p)KW(q,p)^*$, that is,
$$
\G_K(Z) 
=\frac1{2\pi}\int_ZW(q,p)KW(q,p)^*dqdp, \quad Z\in\mathcal B(\R^2),
$$
where $W(q,p) = e^{i\,qp/2}e^{-iqP}e^{ipQ}$ is the Weyl operator.
The Cartesian marginal observables $X\mapsto\G_K(X\times \R)$ and $Y\mapsto\G_K(\R\times Y)$ are, indeed, of the form $\mu^K*\sfq$
and $\nu^K*\sfp$, with the convolving probability measures
$\mu^K(X)= p^{\sfq}_{\Pi K\Pi^*}(X)$ and $\nu^K(Y)=  p^{\sfp}_{\Pi K\Pi^*}(Y)$, where $\Pi=\Pi^*=U_{\pm\pi}$ is the parity operator $(\Pi\psi)(x)=\psi(-x)$.
For any (bounded) operator $A$ we will use the following short notation:
$$
A_\theta:=U_\theta A U_\theta^*.
$$
Especially, $\Pi K\Pi^*=K_\pi$.
Consistently with the notation $\rho^{\sfq}$, the densities of $\mu^K$ and $\nu^K$ may be written as $K_\pi^{\sfq}=
(\Pi K\Pi^*)^{\sfq}=
K^{\Pi^*\sfq\Pi}=K^{\sfq_\pi}$ and $K_\pi^{\sfp}=K^{\sfp_\pi}$.

One may  also determine the marginal observables of $\G_K$ with respect to a rotated orthonormal frame $\{{\bf e}_1(\theta),{\bf e}_2(\theta)\}$ of $\R^2$ where
$$
{\bf e}_1(\theta):=(\cos\theta,\sin\theta), \hspace{1cm}{\bf e}_2(\theta):=(-\sin\theta,\cos\theta)
$$
so that $q_{\theta}{\bf e}_1(\theta)+p_{\theta}{\bf e}_2(\theta)=(q,p)$ with
$$
q_{\theta}=q\cos\theta+p\sin\theta, \hspace{1cm} p_{\theta}=-q\sin\theta+p\cos\theta.
$$
Since 
$U_\theta^* W(q,p)U_\theta=W(q\cos\theta+p\sin\theta,-q\sin\theta+p\cos\theta)=W(q_{\theta},p_{\theta})$,
the relevant marginal observables are simply
\begin{eqnarray*}
&&X\mapsto \frac1{2\pi}\int_{X\times\R}W(q,p)KW(q,p)^*dp_{\theta}= (\mu^{K_{-\theta}}*\sfq_\theta)(X),\\
&&Y\mapsto \frac1{2\pi}\int_{\R\times Y}W(q,p)KW(q,p)^*dq_{\theta}= (\nu^{K_{-\theta}}*\sfp_\theta)(Y),
\end{eqnarray*}
that is, they are unsharp quadrature observables, the convolving measures being determined
by the rotated generating operator $K_{-\theta}$. For any state $\rho$, the density of  $\mu^{K_{-\theta}}*\sfq_\theta$, say, is the
convolution $K_{\pi-\theta}^{\sfq}*\rho^{\sfq_\theta}$
of the densities $K_{\pi-\theta}^{\sfq}=(\Pi K_{-\theta}\Pi^*)^{\sfq}$ and $\rho^{\sfq_\theta}=\rho_{-\theta}^{\sfq}$.

The "$\theta$-marginal observables"   $\mu^{K_\theta}*\sfq_\theta$ of $\G_K$ constitute the Radon transform $\G_K$. Indeed,
for any state $\rho$, let $g^{\rho}_K(q,p):=\tr{\rho W(q,p)KW(q,p)^*}$ denote the density of $\G_K$ in the state $\rho$.
The Radon transform of  $g^\rho_K$ is defined as
$$
({\mathsf R} g^\rho_K)(\theta,q_{\theta}):=\int_\R g^\rho_K\underbrace{(q_{\theta}\cos\theta-p_{\theta}\sin\theta,\,q_{\theta}\sin\theta+p_{\theta}\cos\theta)}_{=\,(q,p)}dp_{\theta}
$$
This shows that for any $\theta\in[0,2\pi)$, the function $q_{\theta}\mapsto ({\mathsf R} g^\rho_K)(\theta,q_{\theta})$
is  ($2\pi$ times) the density $K_{\pi-\theta}^{\sfq}*\rho^{\sfq_\theta}$
of the probability measure 
$\mu^{K_{-\theta}}*p^{\sfq_\theta}_\rho$ of the unsharp rotated quadrature observable $\mu^{K_{-\theta}}*\sfq_\theta$ 
in the state $\rho$. 

Recall that,
for fixed $\theta$ and $q_{\theta}$, the image $\ell(\theta,q_{\theta})$ of
$$
\R\ni p_{\theta}\mapsto(q_{\theta}\cos\theta-p_{\theta}\sin\theta,\,q_{\theta}\sin\theta+p_{\theta}\cos\theta)\in\R^2
$$
is a line on the plane $\R^2$. It goes through a point $(q_{\theta}\cos\theta,q_{\theta}\sin\theta)=q_{\theta} {\bf e}_1(\theta)$ and its direction unit vector is ${\bf e}_2(\theta)$.
Hence, in the above definition of $\mathsf R$, the integral is a line integral over $\ell(\theta,q_{\theta})=q_{\theta} {\bf e}_1(\theta)+\R {\bf e}_2(\theta)$ which is perpendicular to the vector $q_{\theta} {\bf e}_1(\theta)$.

Fixing $\theta$, let $X\in\mathcal B(\R)$, and define 
$$
Z(\theta,X):=\{q_{\theta} {\bf e}_1(\theta)+p_{\theta}{\bf e}_2(\theta)\in\R^2\,|\,q_{\theta}\in X,\,p_{\theta}\in\R\}=\bigcup_{q_{\theta}\in X}\ell(\theta,q_{\theta})
\in\mathcal B(\R^2).
$$
Then
\begin{eqnarray*}
\tr{\rho\G_K(Z(\theta,X))}&=&\frac{1}{2\pi}\int_{Z(\theta,X)} g_K^\rho(q,p)dqdp=\frac{1}{2\pi}\int_{X} ({\mathsf R} g_K^\rho)(\theta,q_{\theta}) dq_{\theta} \\
&=&\int_X (K_{\pi-\theta}^{\sfq}*\rho^{\sfq_\theta})(q_{\theta}) dq_{\theta}.
\end{eqnarray*}
For example, in the case of the number state $K=|h_n\rangle\langle h_n|$, one gets 
$$
\G_{|h_n\rangle\langle h_n|}(Z(\theta,X))=\frac1{2^n n! \sqrt\pi}\int_X \int_\R [H_n(x-q_{\theta})]^2 e^{-(x-q_{\theta})^2} \sfq_\theta(dx)dq_{\theta}
$$
(where $H_n$ is the $n$th Hermite polynomial)
and especially
$$
\G_{|h_0\rangle\langle h_0|}(Z(\theta,X))=\frac1{\sqrt\pi}\int_X \int_\R e^{-(x-q_{\theta})^2} \sfq_\theta(dx)dq_{\theta}.
$$

Let $\mathsf W_\rho$ be the Wigner function of the state $\rho$, that is, $\mathsf W_\rho(q,p)=\frac 1\pi\tr{\rho W(q,p)\Pi W(q,p)^*}$.
As well-known, the Radon transform of the (integrable) Wigner function of a state is the rotated quadrature distribution of this state, that is,
$({\mathsf R}\mathsf W_\rho)(\theta,x)=\rho^{\sfq_\theta}(x)$, see, for instance, \cite{Leonhardt97}.
Therefore, the density $x\mapsto ({\mathsf R} g^\rho_K)(\theta,x)$
is a smearing of the density $x\mapsto({\mathsf R}\mathsf W_\rho)(\theta,x)$ with the density $K_{\pi-\theta}^{\sfq}$.
This observation bring us to the tomography {\small POM} $\Eh$.

\section{The pairs $(Q,Q_\theta)$ and the tomography pom $\Eh$}\label{tomography}
For any state $\rho$,
the random sampling of the distributions $\rho^{\sfq_\theta}$, $\theta\in[0,2\pi)$, determines a probability bimeasure 
$(\Theta,X)\mapsto\int_\Theta\int_X\rho^{\sfq_\theta}(x)\frac{dx d\theta}{2\pi}$, which, when taken all together,
determine  the  {\it tomography observable}, 
$$
\Eh(\Theta\times X):=\frac{1}{2\pi}\int_\Theta \sfq_\theta(X)d\theta,
$$
studied extensively, for instance, in \cite{Albini}. In particular, $\Eh$ is informationally complete.
Its marginal observables are  $\Theta\mapsto \frac{1}{2\pi}\int_\Theta d\theta\,I$ and
$$
X\mapsto \frac{1}{2\pi}\int_0^{2\pi} \sfq_\theta(X)d\theta=\sum_{n=0}^\infty\int_X [h_n(x)]^2 dx \, |{h_n}\rangle\langle{h_n}|
$$ 
since $\sfq_\theta(X)=\sum_{n,m=0}^\infty e^{i(n-m)\theta}\int_X h_n(x)h_m(x) dx \, |{h_n}\rangle\langle{h_m}|$.

The statistics of a phase space observable $\G_K$ can be obtained from the statistics of the tomography observable $\Eh$ in terms of 
a generalized Markov kernel, at least whenever the generating operator $K$ is smooth, that is,  the integral kernel of $K$ belongs to the Schwartz space of $\R^2$.
 Indeed, in that case  one may write for any compact $Z\subset\R^2$ 
$$
\G_K(Z)=\int_0^{2\pi}\int_\R\left[\frac1{2\pi}\int_Z M_{q,p}^K(\theta,x)dqdp\right]d\Eh(\theta,x)
$$
where $M_{q,p}^K(\theta,x)$ is a smooth function with respect to all variables \cite[sect.\ 3.2]{P2009}.
(We call this function a generalized Markov kernel since it is not necessarily positive.)
For instance, if $K=|h_n\rangle\langle h_n|$, the function $M_{q,p}^K(\theta,x)$ takes the form \cite[eq.\ (3.7), Thm 1]{P2009}
\begin{eqnarray*}
M_{q,p}^{|h_n\rangle\langle h_n|}(\theta,x) &=& M_{0,0}^{|h_n\rangle\langle h_n|}(0,x-q_{\theta}) \\
&=&
\sum_{u=0}^n{n\choose u}\frac{2^{1-u}}{u!}\frac{\partial^{2u+1}}{\partial x^{2u+1}}\left[e^{-(x-q_\theta)^2}\int_0^{x-q_\theta}e^{y^2}dy\right] \\
&=&
\sum_{k=n}^\infty {k\choose n}\frac{(-1)^{k-n} k!}{2^k(2k)!}H_{2k}(x-q_\theta).
\end{eqnarray*}
For any smooth trace-class operators $K$ and $\rho$ we have \cite[eq.\ (3.3)]{P2009}
$$
g_K^\rho(q,p)=\int_\R\int_0^{2\pi} M^K_{q,p}(\theta',x)\rho^{\sfq_{\theta'}}(x)\frac{d\theta'}{2\pi} d x
$$
so that
$$
(K_{\pi-\theta}^{\sfq}*\rho^{\sfq_\theta})(q_\theta)=
\int_\R K^{\sfq_{\theta}}(x-q_\theta)\rho^{\sfq_{\theta}}(x)dx
=\frac1{(2\pi)^2}\int_\R\int_\R\int_0^{2\pi} M^K_{q,p}(\theta',x)\rho^{\sfq_{\theta'}}(x)d\theta' d x d p_\theta.
$$

The phase space observable $\G_K$, 
with  $K$ commuting with $N$, 
that is, $K=\sum_{n=0}^\infty w_nP[h_n]$, $0\leq w_n\leq 1$, $\sum_{n=0}^\infty w_n=1$, 
is of special interest since then the angle marginal observable (with respect to the polar coordinates) of $G_K$
is phase shift covariant, that is, a phase observable  \cite[thm 4.1]{LP1999}.
Clearly, in that case
$K^\sfq_\theta=K^\sfq$ for all $\theta\in\R$.
If $K$ is smooth and commutes with $N$, then 
$M_{0,0}^K(\theta,x)=M_{0,0}^K(0,x)$ for all $\theta,\,x\in\R$.

\section{Measuring the pairs $(Q,Q_\theta)$}\label{measuring}
As already pointed out, there is no {\small {\small POM}} having $\sfq$ and $\sfq_\theta$ as its marginal observables.
These observables do not have any joint measurements. Apart from that there are single  measurement schemes which allow one to determine 
both the $\sfq$ and the $\sfq_\theta$ -distributions
$\rho^{\sfq}$ and $\rho^{\sfq_\theta}$  for a large class of states $\rho$. 

To illustrate this possibility, consider a sequential combination of the standard von Neumann measurements of first $\sfq$ and then $\sfq_\theta$,
as described, e.g.\ in \cite[III.2.6]{QTM}. Such a sequential measurement defines a unique phase space observable $\G$ \cite{Davies70,BCL90}. 
Its first marginal observable is an unsharp position  $\mu*\sfq$ defined by the first measurement, whereas its
second marginal observable is an unsharp quadrature  $\nu*\sfq_\theta$ defined by the second measurement under the influence of the first measurement.
The structure of the
convolving measures $\mu$ and $\nu$ depend on the details of the applied measurement schemes, 
in particular, of the initial states of the probe systems. We do not need these details here.

For any state $\rho$ one may determine the moments 
of the marginal distributions $\mu*p^\sfq_\rho$ and $\nu*p^{\sfq_\theta}_\rho$,
of the actual measurement statistics $p^\G_\rho$, 
and they are of the generic form
$$
(\mu*p^\sfq_\rho)[k]= \int_\R x^k d(\mu*p^\sfq_\rho)(x)=\sum_{n=0}^k\binom{k}{n}\mu[k-n]p^\sfq_\rho[n],
$$
where, for instance, $\mu[k]$ denotes the $k^{\rm th}$ moment of $\mu$.
Choosing the initial states of the two probe systems such that  all the moments of $\mu$ and $\nu$ are finite (for instance, choosing the two states
to be Gaussians) and assuming that $\rho=|\psi\rangle\langle\psi|$, with $\psi$ in the linear hull of the Hermite functions, then the actually measured moment sequencies
$((\mu*p^\sfq_\rho)[k])_{k\in\N}$ and $((\nu*p^{\sfq_\theta}_\rho)[k])_{k\in\N}$ can be solved for the sequencies $(p^\sfq_\rho[k])_{k\in\N}$ and 
$(p^{\sfq_\theta}_\rho[k])_{k\in\N}$, which, due to exponential boundedness of the involved probability measures, uniquely determine the distributions $p^\sfq_\rho$ and $p^{\sfq_\theta}_\rho$,
respectively; for technical details, see
 \cite{BKL2008,KLS2009}. 
Note that the distributions  $\rho^\sfq$ and $\rho^{\sfq_\theta}$, with the above choice of $\rho$, suffice to determine the whole observables $\sfq$ and 
$\sfq_\theta$, respectively. 
Note also that the phase space observable $\G$ is not of the form $\G_K$ unless $\theta=\frac\pi{2}$.

To close this section, we recall that the eight-port homodyne detector with a strong local oscillator  is an actual quantum optical implementation of 
a single measurement  scheme which  allows one to determine the distributions $\rho^\sfq$ and $\rho^{\sfq_\theta}$ of
any pair of quadratures $(\sfq,\sfq_\theta)$ for a large class of states \cite{KL2008b}.

\section{Concluding remarks}
We have studied the pairs of quadrature observables $(\sfq,\sfq_\theta)$, showing, in particular, that they share all the familiar coupling properties of the
position-momentum pair $(\sfq,\sfp)$, except their defining property of being a Weyl pair. We have also determined the $\theta$-marginal observables of a covariant
phase space observable $\G_K$, and they turned out to be unharp quadrature observables, the convolving probability measure being determined by the rotated
generating operator $K_{-\theta}$.
These marginal observables 
$\mu^{K_{-\theta}}*\sfq_\theta$
constitute the Radon transform of the phase space observable $\G_K$. Since the Radon transform ${\mathsf R}\mathsf W_\rho$ of the
Wigner function  $\mathsf W_\rho$
of a state $\rho$  gives the quadrature distributions $\rho^{\sfq_\theta}$
in that state, we also exhibited the construction of the phase space observable $\G_K$ in terms of the
tomography observable $\Eh$ defined as a random sampling of the quadrature distributions
$\rho^{\sfq_\theta}$, $\theta\in[0,2\pi)$.
We also showed that in spite of the fact that the quadrature observables $\sfq$
and $\sfq_\theta$ are complementary observables in the sense that they have no joint measurements, it is, anyway, possible to measure the two observables together with a single measurement
scheme, using, for instance, the statistical method of moments.

\end{document}